\documentstyle[12pt]{article}
\begin{document}

\begin{center}
{\LARGE{\bf DOUBLE $\Delta$ PRODUCTION IN THE

\vspace{1cm}

$\gamma d \rightarrow p n \pi^{+} \pi^{-}$ REACTION.}}
\end{center}

\vspace{2cm}

\begin{center}
{\large{J.A. G\'omez Tejedor$^{1}$, E. Oset$^{1,2}$ and H. Toki$^{2}$.}}
\end{center}

\begin{center}
{\small{\it 1. Departamento de F\'{\i}sica Te\'orica and IFIC\\
Centro Mixto Universidad de Valencia - CSIC\\
46100 Burjassot (Valencia), Spain.}}

{\small{\it 2. Department of Physics, Tokyo Metropolitan University.\\
Minami - Ohsawa 1 - 1, Hachioji - shi\\
Tokyo 192 - 03}}
\end{center}

\vspace{3cm}

\begin{abstract}
{\small{We have studied the $\gamma d \rightarrow \Delta^{++} \Delta^{-}$
reaction which requires the collaboration of the two nucleons in deuteron.
By means of a model previously developed for the $\gamma p \rightarrow p
\pi^{+} \pi^{-}$ reaction, the two body exchange currents leading to
double delta creation are derived. A fair agreement is obtained with
a recent experiment, but more precise measurements and the extension to higher
photon energies look advisable in order to see the limits of the present
theoretical approach.}}
\end{abstract}

\newpage

Processes involving necessarily two nucleons in nuclei are particularly
relevant and through them one expects to get insight into nuclear correlations,
together with the reaction mechanisms. This should help us get a unified
picture of reactions like photon or pion absorption and their relationship
to meson exchange currents (MEC), which evidence themselves in a variety of
reactions, like deuteron photodisintegration \cite{1}, $e^{-}$ elastic and
inelastic scattering \cite{2} etc. In this latter reactions MEC appear as
corrections to the leading impulse approximation. The study of reactions where
the MEC, or two body mechanisms, are the leading term provides a good
laboratory to check our theoretical ideas and have these MEC mechanisms under
control for application in a variety of reactions. In this respect the reaction
$\gamma d \rightarrow \Delta \Delta$ offers one such opportunity since
necessarily the two nucleons in the deuteron are involved and excited to
$\Delta$ states. The recent measurement of this reaction \cite{3} offers a
good opportunity to test these ideas, and this is the purpose of the
present work.

Our two body mechanisms for the process are constrained from the reaction
$\gamma p \rightarrow p \pi^{+} \pi^{-}$. Two body MEC are automatically
generated by means of the $\gamma N \rightarrow N \pi^{+} \pi^{-}$
reaction in one nucleon followed by the absorption of
one of the pions in the second nucleon.

A thorough study of the $\gamma p \rightarrow p \pi^{+} \pi^{-}$ reaction
has been done in \cite{4} and reproduces fairly well the experimental cross
sections, invariant mass distributions \cite{5,6}, etc. From this model we
choose the dominant diagrams in which the final $\pi N$ system comes from the
decay of a $\Delta$. The pion absorbed in the second nucleon excites a
$\Delta$ and thus we are led to the diagrams depicted in fig. 1 for the two
$\Delta$ excitation process.

There we depict the
$\Delta^{++} \Delta^{-}$ excitation for $\gamma$ scattering on a pn pair. The
$T=0$ wave function of the deuteron can be written as
$(|pn> - |np>)/\sqrt{2}$. Fig. 1 corresponds
to $\gamma$ scattering with the first
isospin component $|pn>$ of the wave function with the photon being absorbed
either by the $p$ or the $n$. The diagrams corresponding to $\gamma$ scattering
with the second component $|np>$ would be identical to those in fig. 1 by
exchanging the $p$ and $n$, the $\Delta^{++}$ and $\Delta^{-}$ and
replacing the $\pi^{-}$ by a $\pi^{+}$. Since now the final state is
$\Delta^{-} \Delta^{++}$, instead of $\Delta^{++}  \Delta^{-}$ from the first
component, these two sets of diagrams do not interfere, they contribute the
same amount to the cross section and for practical purposes one evaluates the
cross section with the diagrams of fig. 1 ignoring the $1/\sqrt{2}$ factor of
the isospin wave function.
The contribution of the $N^{*}$ (1520) to the $\gamma N \rightarrow N \pi^{+}
\pi^{-}$ reaction was made manifest in \cite{4}.
It interferes with the direct $\Delta$ production process and is responsible
for the resonant like bump appearing in the $\gamma p \rightarrow p \pi^{+}
\pi^{-}$ cross section, which does not appear in the absence of the N$^{*}$
term.

Assuming for the moment the $\Delta$'s as stable particles we obtain the cross
section

$$
\sigma = \frac{M_{d}}{(s - M_{d}^{2})} \int \frac{d^{3}p}{(2 \pi )^{3}}
\int \frac{d^{3}p'}{(2 \pi )^{3}} \frac{M_{\Delta}}{E_{\Delta}(\vec{p})}
\frac{M_{\Delta}}{E_{\Delta} (\vec{p}')} \bar{\Sigma} \Sigma |T|^{2}
$$

\begin{equation}
(2 \pi )^{4} \delta^{4} (k + p_{d} - p - p')
\end{equation}

\noindent
where $M_{d}, M_{\Delta}$ are the deuteron and $\Delta$ masses, $E_{\Delta}
(\vec{p})$ the energy $\sqrt{M_{\Delta}^{2} + \vec{p}^{2}}$ and
$k_{1}, p_{d}, p, p'$ the fourmomenta of the photon, deuteron,
$\Delta^{++}$ and $\Delta^{-}$ respectively.
$T$ is the matrix element for the reaction from the
model of fig. 1 in Mandl and Shaw normalization \cite{7}.

Since the $\Delta's$ are unstable particles we introduce the identity

\begin{equation}
\int dp^{0} \delta (p^{0} - E_{\Delta} (\vec{p})) = 1
\end{equation}

\noindent
but now we make the replacement

$$
\frac{M_{\Delta}}{E_{\Delta}(\vec{p})} \delta (p^{0} - E_{\Delta}(\vec{p}))
\rightarrow - \frac{1}{\pi} Im \frac{M_{\Delta}}{E_{\Delta} (\vec{p})}
\frac{1}{p^{0} - E_{\Delta} (\vec{p}) - i Im \Sigma_{\Delta} \frac{M_{\Delta}}
{E_{\Delta}(\vec{p})}}
$$

\begin{equation}
\approx -\frac{1}{\pi} Im \frac{1}{\sqrt{s_{1}} - M_{\Delta} +
\frac{i \Gamma (s_{1})}{2}}
\end{equation}

\noindent
and the same expressions of eqs. (2) and (3) in the $p'$ variable,
with $s_{1} = p^{02} - \vec{p}^{2}$, and $\Gamma (s_{1})$ the width of the
$\Delta$ at rest with invariant mass $\sqrt{s_{1}}$, \cite{4}

With the modifications introduced by eqs. (2),(3) we write the final
expression for the cross section

$$
\sigma = \frac{4 M_{d}}{(s - M_{d}^{2})} \int \frac{d^{4}p}{(2 \pi )^{4}}
Im \frac{1}{\sqrt{s_{1}} - M_{\Delta} + \frac{i \Gamma (s_{1})}{2}} Im
\frac{1}{\sqrt{s_{2}} - M_{\Delta} + i \frac{\Gamma (s_{2})}{2}}
$$

\begin{equation}
\bar{\Sigma} \Sigma |T|^{2}
\end{equation}

\noindent
with $p' = k + p_{d} - p$ from momentum conservation and $s_{2} = p'^{02} -
\vec{p}\, '^{2}$.

The nuclear matrix element $T$ is given, with the definition for the
intermediate momentum q given in fig. 1, by

\begin{equation}
T = \int \frac{d^{3}q}{(2 \pi )^{3}} \tilde{\phi} (\vec{q} +
\frac{\vec{p} - \vec{p}\, ' + \vec{k}}{2}) \tilde{T}
\end{equation}

\noindent
where $\tilde{\phi} (k)$ is the deuteron relative wave function in momentum
space

\begin{equation}
\tilde{\phi} (k) = \int d^{3} x e^{i \vec{k} \vec{x}} \phi (\vec{x}) \; ;
\int d^{3}x |\phi (\vec{x})|^{2} = 1
\end{equation}

\noindent
and $\phi (\vec{x})$ is the deuteron relative wave function in coordinate
space which we take
from ref. \cite{8} keeping only the $s$-wave part. The deuteron
wave function has been separated into CM and relative coordinates and the CM
wave function has led to the conservation of momentum after integration over
the CM variables. The particular choice of the internal variables in fig. 1 is
done such that the wave function appears with the same argument in
all terms as shown in eq. (5).

The two body matrix $\tilde{T}$ corresponding to diagrams a) b) c) of fig. 1
is given by (see appendix of ref. \cite{4} for the effective Lagrangians and
Feynman rules)

$$
- i \tilde{T} = e(\frac{f^{*}}{\mu}) ^{2} F (q)^{2} \{
\vec{S}_{1}^{\dagger} \cdot \vec{\epsilon} \, \vec{S}_{2}^{\dagger} \cdot
(\vec{k} + \vec{q}) \frac{i}{(k + q)^{2} - \mu^{2} + i \epsilon}
$$

$$
+ \vec{S}_{1}^{\dagger} \cdot \vec{q} \, \vec{S}_{2}^{\dagger} \cdot
\vec{\epsilon} \frac{i}{q^{2} - \mu^{2} + i \epsilon}
$$

\begin{equation}
+ i 2 \vec{q} \cdot \vec{\epsilon} \, \vec{S}_{1}^{\dagger} \cdot \vec{q} \,
\vec{S}_{2}^{\dagger} \cdot
(\vec{k} + \vec{q}) \frac{i}{q^{2} - \mu^{2} + i \epsilon} \;
\frac{i}{(q+k)^{2} - \mu^{2} + i \epsilon}
\end{equation}

\noindent
where the variable $q^{0}$ is given by $q^{0}= E_{d}/2 -p^{0}$,
$e$ is the electron charge, $\vec{\epsilon}$ the photon polarization vector
(in Coulomb gauge, $\epsilon^{0}=0, \quad \vec{\epsilon} \, \cdot
\vec{k}= 0), \;
\vec{S}$ the transition spin matrix from 1/2 to 3/2 and F(q) a monopole form
factor with $\Lambda = 1.3 \, GeV$. Although quark models \cite{9}, or model
calculations of the $NN$ interaction using correlated two pion exchange
\cite{10}, suggest smaller values of $\Lambda$, our input is related to the
one boson exchange model and for consistency we must use the form factor
determined with these models \cite{11}. However, we have checked the
sensitivity of our results to the form factor. By using $\Lambda = 1000 \; GeV$
we find that the cross section is decreased by less than $5\%$ in the whole
energy range that we study.

The integral of eq. (5) with the amplitude of eq. (7) contains the pion
propagator and both the principal part and the pion pole parts are evaluated.

So far we have avoided to include the $N^{*} (1520)$ contribution because this
can be done in an easy way. Once more we refer the reader to ref. \cite{4} for
the effective Lagrangians and couplings. The terms d) e) of fig. 1 have
exactly the same spin structure as the terms a) and c) in that figure and their
incorporation into the amplitude is done, as shown in section 3 of ref.
\cite{4}, by means of the substitution in terms a) and c) of fig. 1
(first and second terms in eq. (7) )

\begin{equation}
e \frac{f^{*}}{\mu} \vec{S}^{\dagger} \cdot \vec{\epsilon}
\rightarrow \{ e \frac{f^{*}}
{\mu} - \frac{\tilde{f}_{N'^{*} \Delta \pi}}{\sqrt{s} - M_{N'^{*}} + i
\frac{\Gamma^{*}}{2}(s)} (\tilde{g}_{\gamma} - \tilde{g}_{\sigma} ) \}
\vec{S}^{\dagger} \cdot \vec{\epsilon}
\end{equation}

\noindent
with $\sqrt{s}$ the invariant $N^{*}$ mass in the diagrams, $\Gamma^{*} (s)$
the $N^{*}$ decay width \cite{4}. We take the values for the coupling constants

$$
\tilde{f}_{N'^{*} \Delta \pi} = 0.677
$$

$$
\tilde{g}_{\gamma} = 0.108 \; ; \; \tilde{g}_{\sigma} = -0.049 \, for
\, N^{*} \rightarrow \gamma p
$$

\begin{equation}
\tilde{g}_{\gamma} = -0.129; \; \tilde{g}_{\sigma} = 0.00731 \, for \, N^{*}
\rightarrow \gamma n
\end{equation}

\noindent
which are required to reproduce the helicity amplitudes in the $N^{*} (1520)
\rightarrow N \gamma$ decay and the decay of $N^{*} (1520)$ into the
$\Delta \pi$ system \cite{4}.

We show the results in fig. 2 and compare them to the available experimental
results \cite{3}. As we can see, the results agree rather well with the data
in the low energy range, but in the highest measured point, the calculated
cross section is smaller then the experimental one. We should point out here
that the dominant terms in figs. 1 are those involving the Kroll Ruderman term,
1a, 1c. Diagram 1b gives a smaller contribution and when added to the dominant
terms 1a + 1c it changes the cross section only at the level of $10\%$
(it decreases $\sigma$ below $E_{\gamma}= 740 MeV$ and increases it above this
energy).
We also show in the
figure the effects of considering the $N^{*}(1520)$ resonance. There is a
constructive interference below $E_{\gamma} \simeq 820 MeV$ and a destructive
interference above this energy, due to the change of sign of the real part of
the $N^{*}$ propagator. A similar thing happens in the case of the
$\gamma p \rightarrow p \pi^{+} \pi^{-}$ reaction and the interference
is responsible for the bump in the cross section shown by the experiment.
In spite of the large experimental errors \cite{3} we still can see that
the cross section in the experiment of fig. 2 is better reproduced by the
inclusion of the $N^{*}$ term. In ref. \cite{4} the coupling $N^{\ast} (1520)
\rightarrow \Delta \pi$ is taken as a constant. At high energies some small
momentum dependent components of this coupling could become more important.
However, we have used a quark model picture to evaluate this vertex and have
found that the differences between
the results with the constant coupling and the ones of the
quark model, providing extra momentum dependent terms, are not significant in
the $\gamma p \rightarrow p \pi^{+} \pi^{-}$ reaction \cite{12}.

In fig. 2 we also show our results at higher
energies. As we can observe,
the cross section stops increasing around $E_{\gamma}= 1060 MeV$ and starts
going down smoothly from there on. However, one should also be aware that
other MEC terms generated from the model of ref. \cite{4} and leading to
$2 \Delta$ excitation could become relevant in the region of $E_{\gamma} >
800 MeV$. We should also note that our model for $\gamma p \rightarrow
\pi^{+} \pi^{-} p$ starts having discrepancies with the data at energies
$E_{\gamma}> 800 \; MeV$.
Furthermore, as the photon energy increases one is
picking up larger momentum components of the deuteron (see the argument of
$\tilde{\phi}$ in eq. (5) and the $d$-wave component could also play a role.
For all these reasons our results for $E_{\gamma} > 800 \; MeV$ have larger
uncertainties as the photon energy increases.

In summary we have studied the reaction
$\gamma d \rightarrow \Delta^{++} \Delta^{-}$ which is a genuine two body
process. The mechanisms for the reaction were obtained by
studying previously the $\gamma N \rightarrow N \pi^{+} \pi^{-}$ reaction
and choosing the diagrams where one nucleon and a pion emerge in a $\Delta$
resonant state. The second pion  was absorbed by the second nucleon exciting
also a $\Delta$.
The two body mechanisms generated in this way reproduce fairly well
the experimental cross section at low energies but the results seem to be
lower than experiment at higher energies, although the experimental data
show strong oscillations there. The common
features of the approach used here with current microscopic approaches for
photon absorption in nuclei gives extra support to these approaches
and also strengthens our confidence in the use of two body meson
exchange currents in other reactions.

In order to know the limits of the
present method, more accurate data and extension of the measurements to
higher energies would be most welcome.

\vspace{8cm}

\noindent
We would like to acknowledge partial support from the CICYT contract no.
AEN 93 - 1205. One of us E.O. wishes to acknowledge the hospitality of Tokyo
Metropolitan University and support from the Japan Society for the promotion
of Science. JAGT wishes to acknowledge support from IVEI.

\newpage

figure captions.

fig. 1.- Terms considered in our model for the $\gamma d \rightarrow
\Delta^{++} \Delta^{-}$ reaction.

fig. 2.- Results of the model compared to the data of ref. \cite{3}. Dashed
line, omitting the $N^{*}$ terms (fig. 3d, 3e). Solid line, results including
all terms of the model of fig. 1.


\newpage

\begin{thebibliography}{99}
\bibitem{1} H.J. Weber and H. Arenh\"{o}vel, Phys. Reports 36 (1978) 277
\bibitem{2} B. Frois, Prog. Part. Nucl. Phys. 13 (1985) 117;\\
J. E. Amaro, C. Garc\'{\i}a-Recio and A.M. Lallena, Nucl. Phys. in print;
S. Moraghe, J.A. Amaro, C. Garc\'{\i}a-Recio and A.M. Lallena, Univ. Granada
preprint
\bibitem{3} M. Asai et al., Z. Phys. A344 (1993) 335
\bibitem{4} J.A. G\'omez-Tejedor and E. Oset, Nucl. Phys. A571 (1994) 667
\bibitem{5} Aachen - Berlin - Bonn - Hamburg - Heidelberg - M\"{u}nchen
collaboration, Phys. Rev. 175 (1968) 1669
\bibitem{6} G. Gianella et al., Nuovo Cimento LXIII A (1969) 892
\bibitem{7} F. Mandl and G. Shaw, Quantum Field Theory, John Wiley, 1984
\bibitem{8} M. Lacombe et al., Phys. Lett. 101 B (1981) 139
\bibitem{9} R. Tegen and W. Weise, Z. Phys. A314 (1983) 357
\bibitem{10} H.C. Kim, J.W. Durso and K. Holinde, Phys. Rev. C49 (1994) 2355
\bibitem{11} R. Machleidt, K. Holinde and Ch. Elster, Phys. Reports
149 (1987) 1
\bibitem{12} F. Cano, J.A. G\'omez-Tejedor, P. Gonz\'alez and E. Oset, to be
published.
\end{thebibliography}
\end{document}